\documentclass[aps,pre,twocolumn,groupedaddress,superscriptaddress,showpacs]{revtex4-1}

\usepackage{graphicx}
\usepackage{color}
\usepackage{amsmath}
\usepackage{amsfonts}
\usepackage{amssymb}
\usepackage{dcolumn}
\usepackage{hyperref}
\hypersetup{colorlinks=true,urlcolor=blue,linkcolor=blue,citecolor=blue}

\begin{document}
\title{Retention capacity of correlated surfaces}
\author{K. J. Schrenk}
\email{jschrenk@ethz.ch}
\affiliation{Computational Physics for Engineering Materials, IfB, ETH Zurich, Wolfgang-Pauli-Strasse 27, CH-8093 Zurich, Switzerland}
\author{N. A. M. Ara\'ujo}
\email{nuno@ethz.ch}
\affiliation{Computational Physics for Engineering Materials, IfB, ETH Zurich, Wolfgang-Pauli-Strasse 27, CH-8093 Zurich, Switzerland}
\author{R. M. Ziff}
\email{rziff@umich.edu}
\affiliation{Center for the Study of Complex Systems and Department of Chemical Engineering, University of Michigan, Ann Arbor, Michigan 48109-2136 USA}
\author{H. J. Herrmann}
\email{hans@ifb.baug.ethz.ch}
\affiliation{Computational Physics for Engineering Materials, IfB, ETH Zurich, Wolfgang-Pauli-Strasse 27, CH-8093 Zurich, Switzerland}
\affiliation{Departamento de F\'{\i}sica, Universidade Federal do Cear\'a, 60451-970 Fortaleza, Cear\'a, Brazil}
\begin{abstract}
We extend the water retention model [C. L. Knecht \emph{et al.}, Phys. Rev. Lett. {\bf 108}, 045703 (2012)] to correlated random surfaces.
We find that the retention capacity of discrete random landscapes is strongly affected by spatial correlations among the heights.
This phenomenon is related to the emergence of power-law scaling in the lake volume distribution.
We also solve the uncorrelated case exactly for a small lattice and present bounds on the retention of uncorrelated landscapes.
\end{abstract}
\maketitle
\section{\label{sec::intro}Introduction}
\begin{figure}
	\includegraphics[width=\columnwidth]{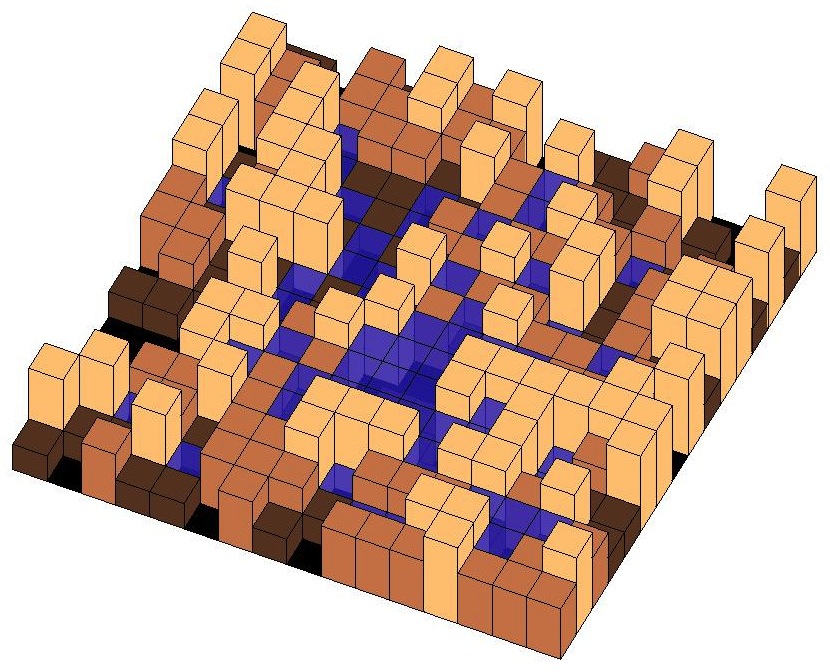}
	\caption{
		\label{fig::L_16_uniform_canonical_4_levels_retention_72}
		(Color online)
		Snapshot of the water retention model on a square lattice of size ${16\times16}$, with number of levels ${n=4}$.
		The colors of the cubes indicate the landscape height, which varies from zero (black) to ${n-1=3}$ (light orange, light gray).
		The retention capacity of this landscape is $72$ (lattice units).
		Blue (dark gray) transparent squares indicate the surfaces of lakes of retained water.
	}
\end{figure}

\begin{figure}
	\center
	\includegraphics[width=\columnwidth]{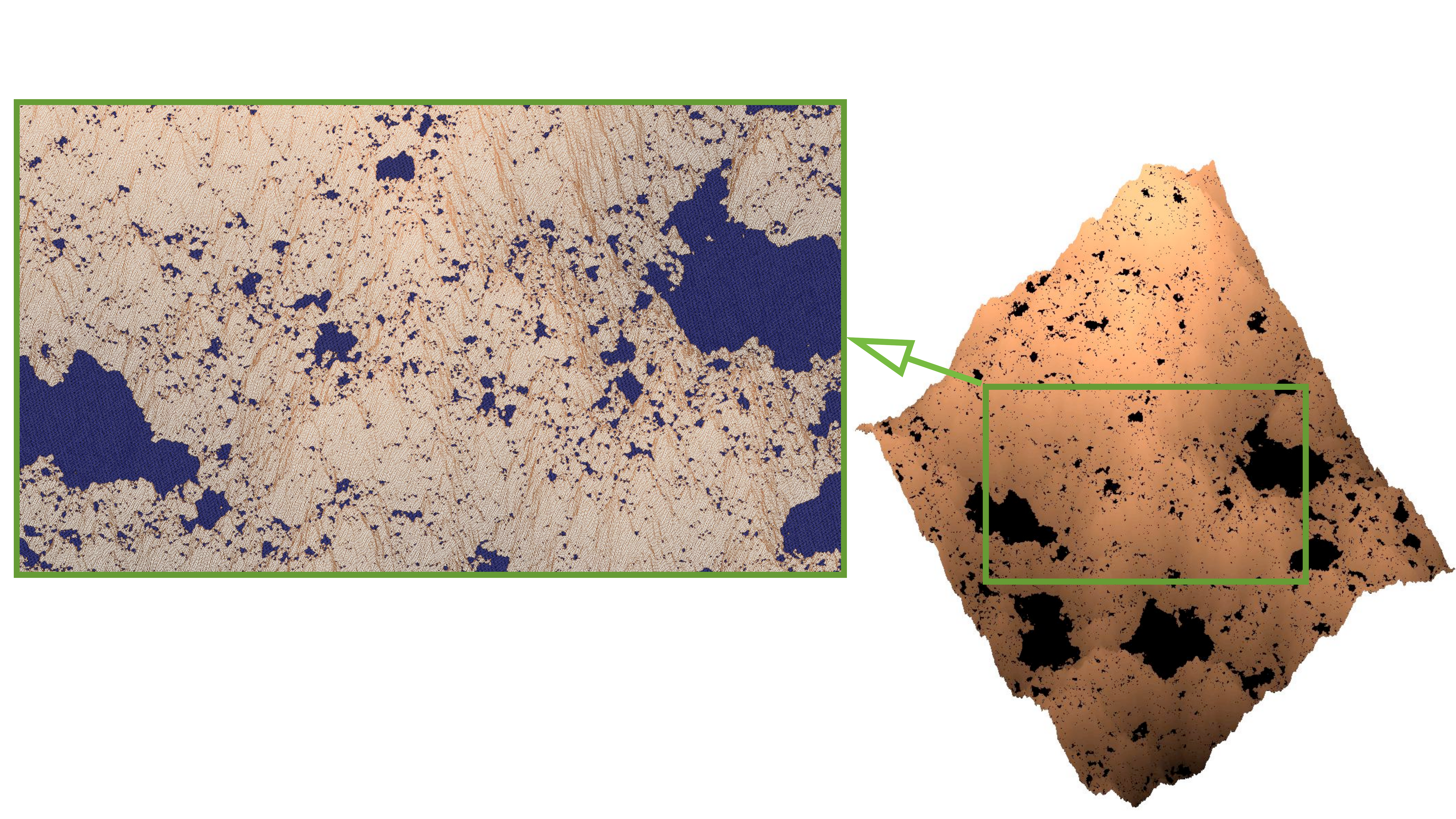}
	\caption{
		\label{fig::water_retention_chapter_continuum_512_H075}
		(Color online)
		Retention model with continuous heights on a lattice of size $512\times512$.
		Surfaces of lakes are indicated in blue.
		The Hurst exponent is ${H=0.75}$, see e.g. Refs.~\cite{Saupe88, Makse96, Ballesteros99b, Kondev00, Ahrens11} for details.
	}
\end{figure}

Consider the discrete landscape in Fig.~\ref{fig::L_16_uniform_canonical_4_levels_retention_72}:
If water rains on this landscape and is allowed to flow out through its open boundaries, one may ask what is the volume of water retained by the landscape, i.e., when the water level of all ponds has reached its maximum.
Recently, Knecht \emph{et al.} investigated this question for random landscapes with uncorrelated heights, coining the term \emph{water retention model} \cite{Knecht11}.
Further studies of the critical behavior of the water retention model have also been reported by Baek and Kim \cite{Baek11}.
An example of a correlated landscape, where all ponds and lakes are at their maximum capacity, is shown in Fig.~\ref{fig::water_retention_chapter_continuum_512_H075}.
Looking at the examples in Figs.~\ref{fig::L_16_uniform_canonical_4_levels_retention_72} and \ref{fig::water_retention_chapter_continuum_512_H075}, one anticipates that the degree of correlation among the heights of the landscapes can drastically impact on their retention behavior.
This subject is studied in the present work.

There has been recently much interest in the properties of random landscapes, with and without spatial correlations \cite{Kondev95, Kondev00, Majumdar06, Schrenk12, Schrenk13b}.
Such landscapes are used on different scales, from deposition phenomena \cite{Barabasi95} to driven movement and transport in random geometries \cite{Araujo02, LeDoussal09}, geomorphology \cite{Xu93, Czirok93, Pastor-Satorras98, Nikora99}, and city growth \cite{Makse98}.
Related concepts have also been generalized to non-Euclidean graphs and their partitioning \cite{Carmi08, Sollich08}.

The water retention capacity is a global property of a random landscape.
A closely related question is how to predict through which part of the boundary the water of spilling ponds will flow out of the landscape.
Ponds spilling to different parts of the boundary are separated by watersheds of the considered landscape \cite{Fehr11, Daryaei12} and share statistical properties of optimum paths and polymers in strongly disordered media \cite{Andrade11}.
For uncorrelated random heights, watersheds are known to be fractals of dimension ${1.2168\pm0.0005}$ \cite{Fehr11c}.
This fractal dimension is affected by long-range correlations in the landscape \cite{Fehr11b}.

Here we report that the retention capacity of random landscapes is strongly affected by long-range height correlations.
This adds to the understanding of this recently introduced model \cite{Knecht11}, because natural landscapes are typically characterized by such correlations \cite{Xu93, Czirok93, Pastor-Satorras98, Nikora99} and the two previous studies of the model dealt only with uncorrelated heights \cite{Knecht11,Baek11}.
We find that the decomposition property of the retention is valid for the correlated case as well.
In addition, we report new derivations of an exact result and some bounds for the uncorrelated case.

For the numerical and analytical treatment of the water retention problem, we use its analogies to percolation \cite{Stauffer94}.
In particular, we use an algorithm based on invasion percolation \cite{Wilkinson83}, and we interpret our results establishing connections to percolation with correlated disorder \cite{Weinrib84, Prakash92, Schmittbuhl93, Mandre11, Schrenk13b} and on rough surfaces \cite{Schmittbuhl93, Kondev95, Olami96}.

The remainder of this article is structured as follows.
Section \ref{sec::classical_wr} discusses the water retention model on uncorrelated random landscapes.
The model is solved exactly for a small lattice in Sec.~\ref{sec::solution_test_L3_wr_chapter}.
The impact of height correlations on the retention is analyzed in Sec.~\ref{sec::correlated_wr}.
Conclusions are drawn in Sec.~\ref{sec::fin_wr_chapter}.
\section{\label{sec::classical_wr}Water retention model}
We recall the original definition of the water retention model, as introduced in Ref.~\cite{Knecht11}, along with some first results.
Consider a square lattice of length $L$, consisting of ${N=L^2}$ sites, with free boundary conditions.
Each site of the lattice, covering a unitary area, can be seen as the base of a square column with a certain height.
The set of all $N$ square columns makes up a discrete landscape (see Fig.~\ref{fig::L_16_uniform_canonical_4_levels_retention_72} and e.g. Refs.~\cite{Kondev00, Knecht11, Schrenk12}).
In the original water retention model, we assume that the heights of the columns are integers in the interval ${[0,n-1]}$, where $n$ is the number of levels, a parameter of the landscape.
For simplicity, all $n$ heights appear with the same probability $1/n$.

To determine the volume of water retained by a given landscape, it is convenient to use the invasion algorithm of Ref.~\cite{Knecht11}.
There, the landscape is invaded, similarly to invasion percolation \cite{Wilkinson83}, starting from its boundaries.
The water level at which a site is first invaded, minus the terrain height at this site, gives the maximum retained volume at that site.

\begin{figure}
	\includegraphics[width=\columnwidth]{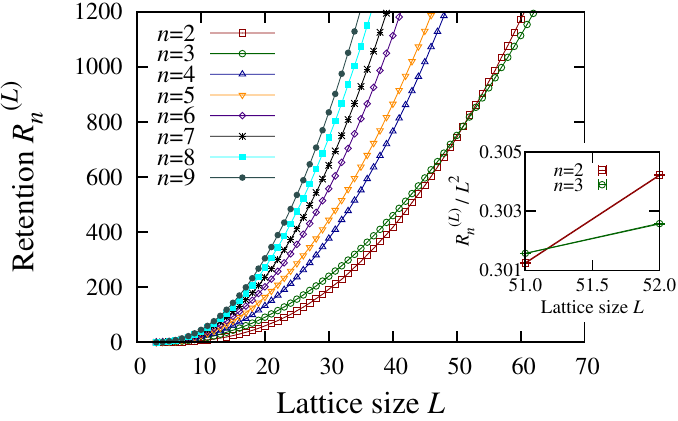}
	\caption{
	(Color online)
	Main plot:
	Retention $R_n^{(L)}$ of an $n$-level system of size $L^2$ as function of $L$.
	Results are averages over at least $10^4$ samples.
	Inset: Data for 2 and 3 levels, showing their crossing behavior.
	Random numbers have been generated with the algorithms of Refs.~\cite{Ziff97, Matsumoto98}.
	\label{fig::wr_FixedSSeveralL}
	}
\end{figure}

\begin{figure}
	\includegraphics[width=\columnwidth]{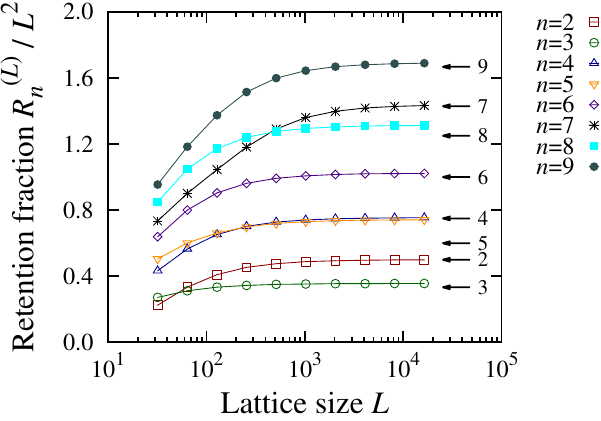}
	\caption{
	(Color online)
	Retention per site ${R_n^{(L)}/L^2}$ of an $n$-level system of size $L^2$ as function of $L$.
	The arrows indicate the lower bounds on ${R_n^{(L)}/L^2}$, for $n$ levels and large $L$, given by Eq.~(\ref{eqn::lower_limit_knecht11_nstar}), with ${p_c=0.59274602}$ \cite{Ziff11b}.
	\label{fig::wr_FixedSSeveralL_PerSite}
	}
\end{figure}

Given an ensemble of landscapes, of size $L$ with $n$ levels, one can define the average volume ${R_n^{(L)}}$ of retained water.
First we consider uncorrelated landscapes as in Ref.~\cite{Knecht11, Baek11}, to illustrate the model.
Figures \ref{fig::wr_FixedSSeveralL} and \ref{fig::wr_FixedSSeveralL_PerSite} show the retention ${R_n^{(L)}}$ for landscapes with equal probability for each of the $n$ heights as function of the lattice size $L$.
In Fig.~\ref{fig::wr_FixedSSeveralL_PerSite}, ${R_n^{(L)}}$ is divided by $L^2$ to show that the retention is proportional to $L^2$ for large lattice sizes $L$.
As reported in Ref.~\cite{Knecht11}, the retention of a landscape of size $L$ is not always monotonically increasing with the number of levels $n$:
For example, one observes in the inset of Fig.~\ref{fig::wr_FixedSSeveralL} that the curves for ${R_2^{(L)}}$ and ${R_3^{(L)}}$ intersect for ${L=51.2\pm0.1}$ \cite{Knecht11}.
This can also be observed in Fig.~\ref{fig::wr_FixedSSeveralL_PerSite}, where for small $L$ the retention grows monotonically in $n$, while for large $L$ this is not always the case, as can be observed for large system sizes for $n=2,3$; $4,5$; and $7,8$ \cite{Knecht11}.
It was argued in Ref.~\cite{Knecht11} that the retention ${R_n^{(L)}}$ of such $n$-level landscapes can be expressed as a sum of terms ${R_2^{(L)}(p)}$ for two-level landscapes with probability $p$ that a site has height $0$ and probability ${1-p}$ that it has height $1$:
\begin{equation}
	\label{eqn::retention_decomposition}
	R_n^{(L)} = \sum_{i=1}^{n-1} R_2^{(L)}(i/n).
\end{equation}
\begin{figure}
	\includegraphics[width=\columnwidth]{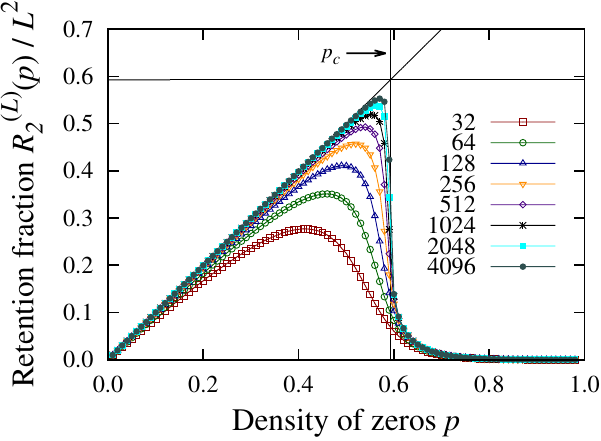}
	\caption{
	(Color online)
	Retention fraction ${R_2^{(L)}/L^2}$ as function of the density of zeros $p$ in the landscape for different $L$.
	The horizontal line indicates the site percolation threshold, ${p_c=0.59274602}$ \cite{Ziff11b}.
	The other straight solid line is the plot of ${f(p)=p}$.
	\label{fig::TwoLevelsVariableProb0}
	}
\end{figure}
\begin{figure}
	\includegraphics[width=\columnwidth]{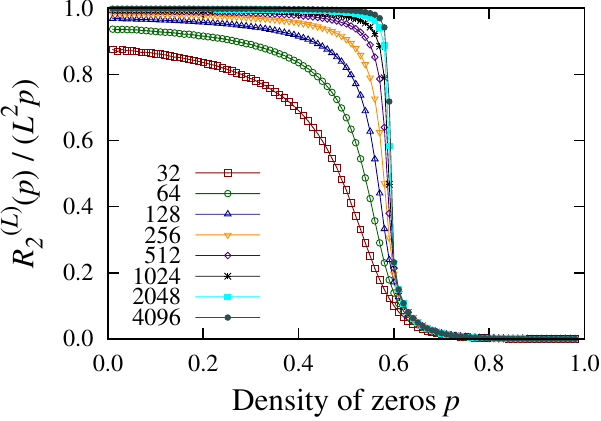}
	\caption{
	(Color online)
	${R_2^{(L)}/(L^2p)}$ as function of the density of zeros $p$ in the landscape.
	\label{fig::TwoLevelsVariableProb0_PerP}
	}
\end{figure}

Figures \ref{fig::TwoLevelsVariableProb0} and \ref{fig::TwoLevelsVariableProb0_PerP} show that, for ${n=2}$ levels, the retention fraction ${R_2^{(L)}(p)/L^2}$, for large $L$, approaches $p$ for $p$ below the site percolation threshold $p_c$ (of the square lattice) and decreases to zero after reaching ${{R_2^{(L)}(p_c)}=p_c}$.
This behavior can be explained by considering
\begin{equation}
	r_2(p) = \lim_{L\to\infty}R_2^{(L)}(p)/L^2,
\end{equation}
the retention fraction of the two-level landscape in the thermodynamic limit, ${L\to\infty}$: Each site of height zero that is not part of the percolating cluster, containing a fraction $P_\infty(p)$ of sites, retains one unit volume of water \cite{Knecht11}, such that
\begin{equation}
	\label{eqn::retention_fraction_uniform}
	r_2(p) = p-P_\infty(p).
\end{equation}

Analyzing the retention in the limits of large and small lattice size $L$ helps understanding why some retention curves show crossings.
We focus first on the behavior of the retention fraction $R_n^{(L)}/L^2$ for large $L$, seen in Fig.~\ref{fig::wr_FixedSSeveralL_PerSite} (the case ${L=3}$ is solved exactly in Appendix~\ref{sec::solution_test_L3_wr_chapter}).
It is useful to find lower and upper bounds on
\begin{equation}
	r_n = \lim_{L\to\infty}R_n^{(L)}/L^2.
\end{equation}
Using Eqs.~(\ref{eqn::retention_decomposition}) and (\ref{eqn::retention_fraction_uniform}), Knecht \emph{et al.} give an approximate expression for $r_n$ \cite{Knecht11}:
This is a lower bound since, for $L\to\infty$, from Eq.~(\ref{eqn::retention_fraction_uniform}), $r_2(p)\geq p\theta(p_c-p)$, such that
\begin{equation}
	\label{eqn::lower_limit_knecht11_nstar}
	r_n
	= \sum_{i=1}^{n-1}r_2\left(\frac{i}{n}\right)
	\geq \sum_{i=1}^{n-1}\frac{i}{n}\theta(p_c-i/n)
	=\frac{n^*(n^*+1)}{2n},
\end{equation}
where $n^*=\left \lfloor{np_c}\right \rfloor$ is the truncated integer part of $np_c$ and $\theta(x)$ is the Heaviside step function, defined as
\begin{equation}
	\theta(x)=
	\left\{
		\begin{array}{ll}
			0 & \mbox{if } x < 0 \\
			1 & \mbox{if } x \geq 0
		\end{array}
	\right.
	.
\end{equation}
Figure~\ref{fig::wr_FixedSSeveralL_PerSite} shows that the bound in Eq.~(\ref{eqn::lower_limit_knecht11_nstar}), indicated by the black arrows on the right-hand side, is consistent with the numerical data and in agreement with the observed crossing behavior for the considered range of $n$.
Without using the decomposition formula in Eq.~(\ref{eqn::retention_decomposition}), one obtains an upper bound on $r_n$ by considering the maximum amount of water that can be retained on a landscape of sufficiently large $L$:
Suppose that $L$ is such that $L^2/n\gtrsim4(L-1)$.
This is enough to ensure that all sites in the boundary of the lattice can be occupied with square columns of the maximum height ${n-1}$.
Since all columns of height smaller than ${n-1}$ are placed in the interior of such a landscape, it retains at most a volume of ${(n-1)L^2/2}$.
Dividing by the number of sites $L^2$ and taking the limit $L\to\infty$ gives:
\begin{equation}
	\label{eqn::upper_limit_on_retention_fraction_r_n}
	r_n\leq(n-1)/2.
\end{equation}
We note that while the lower bound in Eq.~(\ref{eqn::lower_limit_knecht11_nstar}) corresponds to approximating the curve of $r_2(p)$ by a step function which is zero for ${p>p_c}$, the upper bound in Eq.~(\ref{eqn::upper_limit_on_retention_fraction_r_n}) is the same that one would obtain by using ${r_2(p) \leq p}$, for ${0\leq p \leq 1}$ combined with the decomposition formula in Eq.~(\ref{eqn::retention_decomposition}).
Therefore, since ${p_c>1/2}$ \cite{Riordan07, Ziff11b, Jacobsen14}, for ${n=2}$ the upper and lower bounds coincide, and ${r_2=1/2}$ is the exact solution (see Fig.~\ref{fig::wr_FixedSSeveralL_PerSite} and also Fig.~\ref{fig::WaterRetention_prob0_canonical_mean_vs_H}).
Basically, for a two-level system, the clusters of zeros are sub-critical and they become all filled with the water, which sits on exactly half of the sites.
Here, we are assuming $L\to\infty$ so there are no finite-size effects.

We also note that one can use the results in Eq.~(\ref{eqn::lower_limit_knecht11_nstar}), see Refs.~\cite{Knecht11, Baek11}, and Eq.~(\ref{eqn::upper_limit_on_retention_fraction_r_n}) to obtain bounds on the continuum version of the retention model, i.e., the case where the number of levels $n$ becomes infinite.
Dividing the retention per site $r_n$ by $n$, one finds for this limit:
\begin{equation}
	p_c^2/2
	\leq \lim_{n\to\infty} r_n/n \leq
	1/2,
\end{equation}
which is in agreement with the numerical result ${0.1820\pm0.0002}$ \cite{Knecht11} and our simulations (not shown) of the continuum model (we note that $p_c^2/2\approx0.17567$).
\section{\label{sec::correlated_wr}Impact of correlations}
\begin{figure}
	\center
	\begin{tabular}{ll}
		(a) $H=-1$ & (b) $H=-0.5$ \\
		\includegraphics[width=0.48\columnwidth]{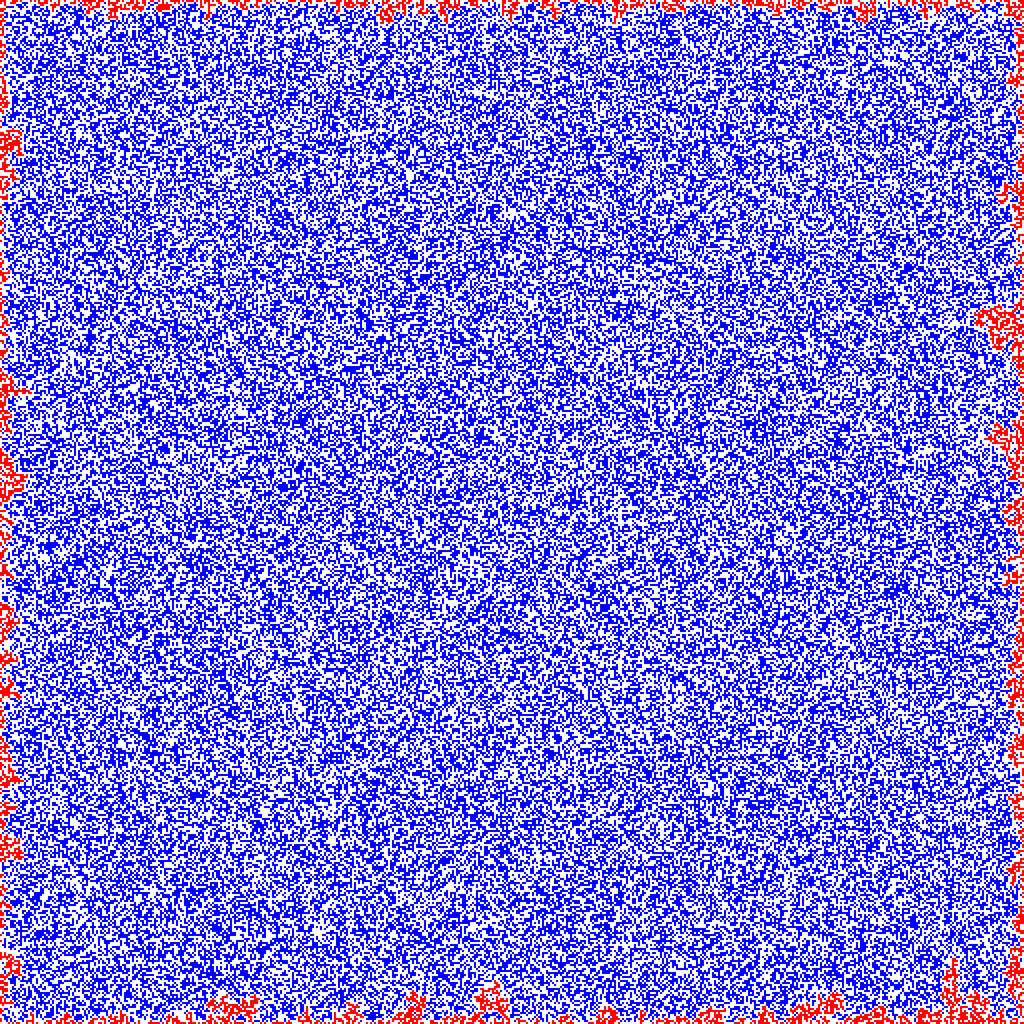}&
		\includegraphics[width=0.48\columnwidth]{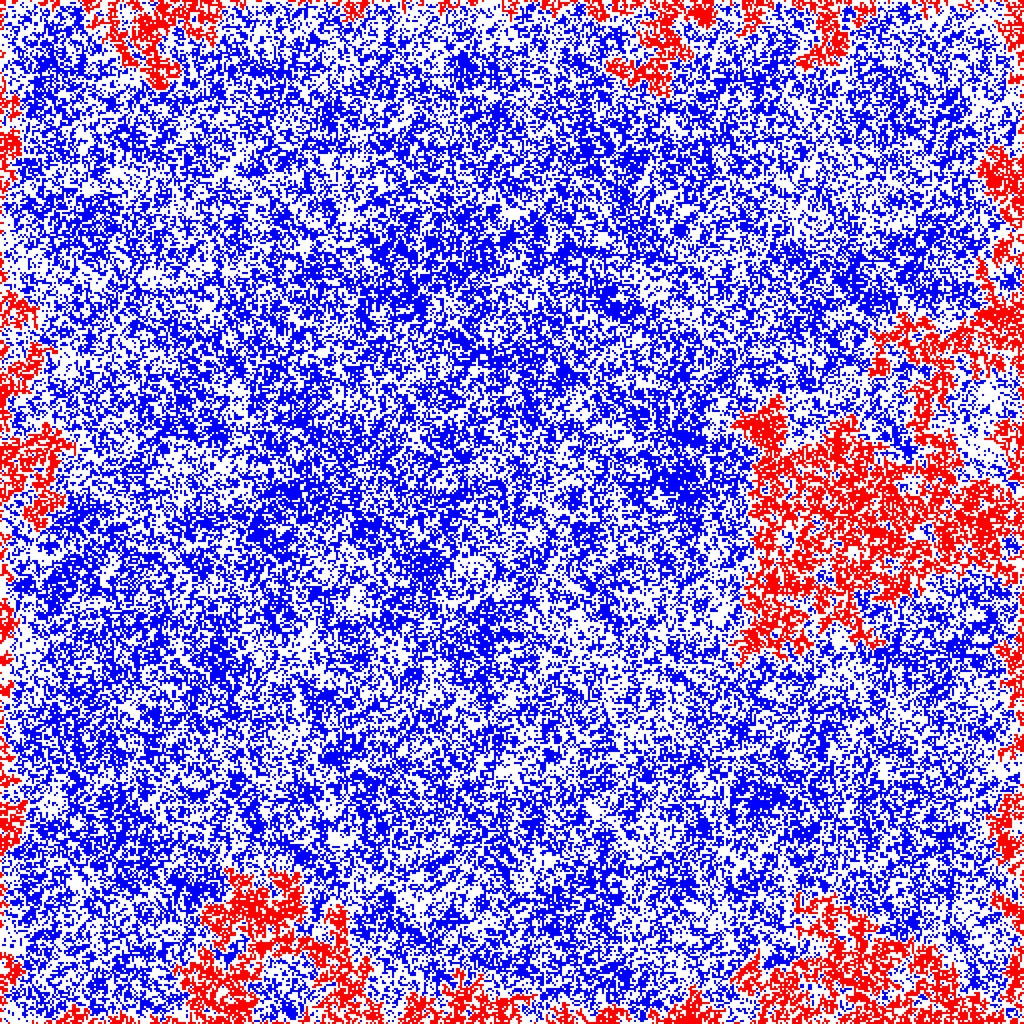}\\
		(c) $H=0$ & (d) $H=0.5$ \\
		\includegraphics[width=0.48\columnwidth]{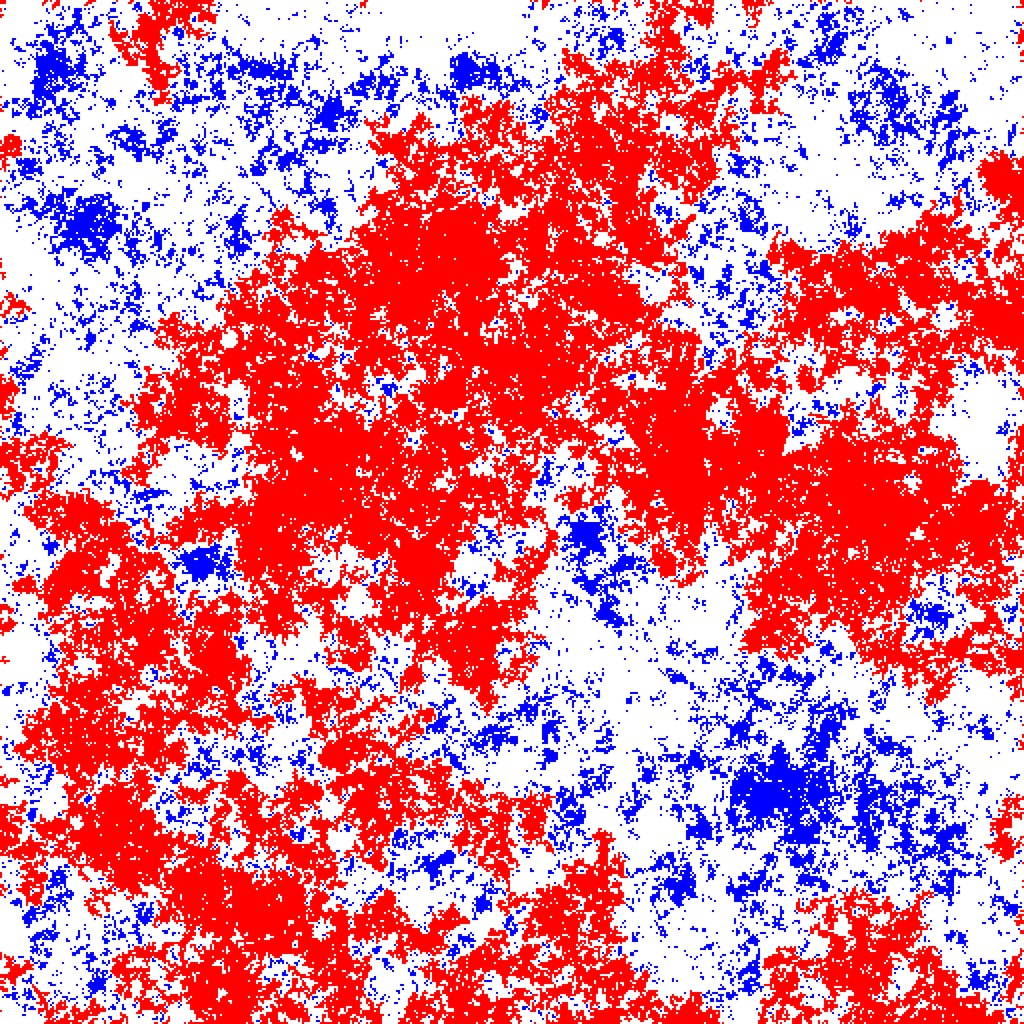}&
		\includegraphics[width=0.48\columnwidth]{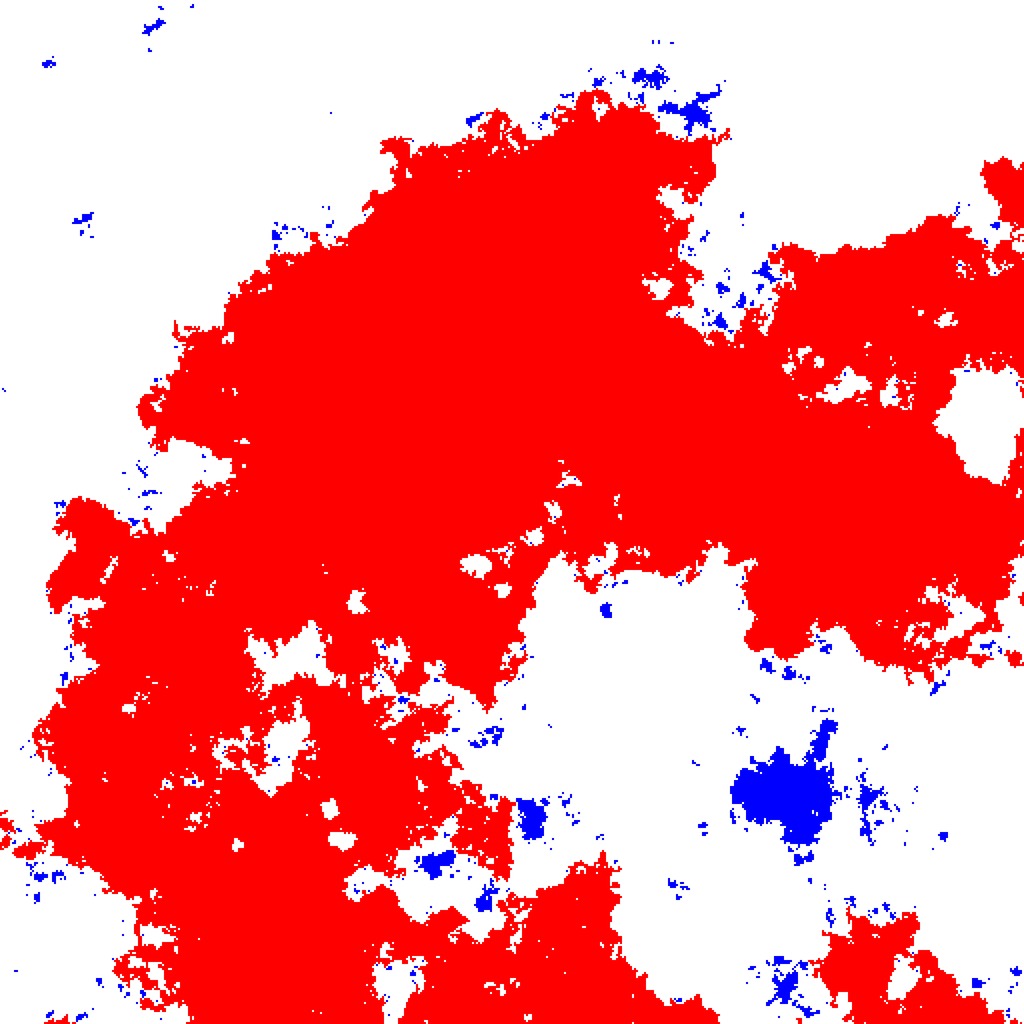}\\
	\end{tabular}
	\caption{
		\label{fig::prob0_canonical_snaps}
		(Color online)
		Snapshots of the water retention model with $n=2$ levels.
		One half of the sites has height $1$, shown as white squares, and the other half has height $0$.
		The sites with height $0$ are colored in blue (dark gray) if they retain water, and they are colored in red (light gray) otherwise (draining to the boundary).
		$H$ increases from (a) to (d) and the lattice size is $L=512$.
		For $H=0$, it is known that the boundaries of the lakes are fractal with dimension $3/2$ \cite{Kondev95, Kondev00, Schrenk13b}, which corresponds to a Gaussian free field or $\text{SLE}_4$ \cite{Schramm09}.
		Furthermore, for $0\leq H \leq1$, it is believed that the boundary fractal dimension is $(3-H)/2$ \cite{Kondev95, Kondev00, Schwartz01}.
	}
\end{figure}
\begin{figure}
	\includegraphics[width=\columnwidth]{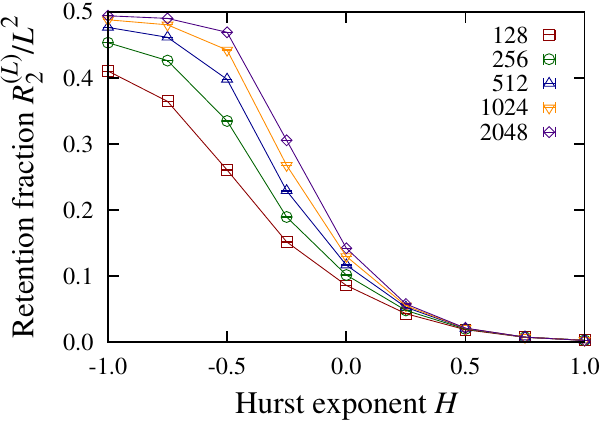}
	\caption{
		\label{fig::WaterRetention_prob0_canonical_mean_vs_H}
		(Color online)
		Retention fraction $R_2^{(L)}/L^2$ for ${n=2}$ levels as function of the Hurst exponent $H$ for different lattice sizes $L$.
	}
\end{figure}
So far, we have considered landscapes with uncorrelated heights.
However, the reason why the landscape in Fig.~\ref{fig::water_retention_chapter_continuum_512_H075} looks natural is that its heights are spatially correlated.
To analyze the impact of long-range correlations of the heights on the retention capacity, it is convenient to consider the canonical version of the model.
There, every height appears exactly $N/n$ times (``canonical''), rather than with probability $1/n$ (``grand canonical'') \cite{Newman00, Newman01, Knecht11, Hu12}.
As discussed in Ref.~\cite{Knecht11}, this does not significantly affect the behavior of the model.
Our correlated landscapes are obtained in the following from the Fourier filtering method with Hurst exponent $H$, corresponding to a power spectrum $S(\mathbf{f})$ of the heights scaling as $S(\mathbf{f})\sim\vert\mathbf{f}\vert^{-2(H+1)}$, for low frequencies $\vert\mathbf{f}\vert$, see e.g. Refs.~\cite{Saupe88, Makse96, Ballesteros99b, Kondev00, Ahrens11}.
Empirically, it has been observed that natural landscapes can be described by $H$ in the range of ${0.3 \lesssim H \lesssim 0.95}$ \cite{Xu93, Czirok93, Pastor-Satorras98, Nikora99}.

To discretize a random landscape into $n$ levels of height $0$ to ${n-1}$, it is convenient to consider the concept of ranked surfaces \cite{Schrenk12}.
The recipe to discretize the continuous landscape is as follows:
First, the ranked surface corresponding to the given landscape is obtained by ranking the sites according to the landscape heights.
Then, one follows the rank of sites, starting from the lowest one, and assigns height $0$ to the first $N/n$ sites.
The next $N/n$ sites in the ranking are assigned height $1$, and so on, until the highest sites have been assigned height $n-1$.
This procedure gives landscapes like the ones in the canonical retention model.
Therefore, discretizing an uncorrelated landscape, e.g. with uniformly and randomly distributed heights, or with $H=-1$, recovers the canonical version, as introduced in Ref.~\cite{Knecht11}.
Here we require that the number of sites can be divided by the number of levels $n$ such that there is no remainder, $L^2\,\text{mod}\,n=0$.
This restricts the possible combinations of $L$ and $n$.

Figure~\ref{fig::prob0_canonical_snaps} shows snapshots of the model for ${n=2}$ levels and different values of $H$.
Retained water is shown in blue while water draining to the boundaries is shown in red.
From these typical configurations, qualitatively speaking, one expects that the mean retention decreases with increasing $H$.
This is confirmed by the data in Fig.~\ref{fig::WaterRetention_prob0_canonical_mean_vs_H}, where the retention fraction $R_2^{(L)}/L^2$ is shown as function of $H$.
By measuring the retention of two-level systems with variable fraction $p$ of sites with height zero $R_2^{(L)}(p)$ (not shown) and comparing this to direct measurements of $R_n^{(L)}$ for $n$-level systems, we confirmed that, within error bars, the decomposition formula in Eq.~(\ref{eqn::retention_decomposition}) is also valid for ${H>-1}$.
\begin{figure}
	\includegraphics[width=\columnwidth]{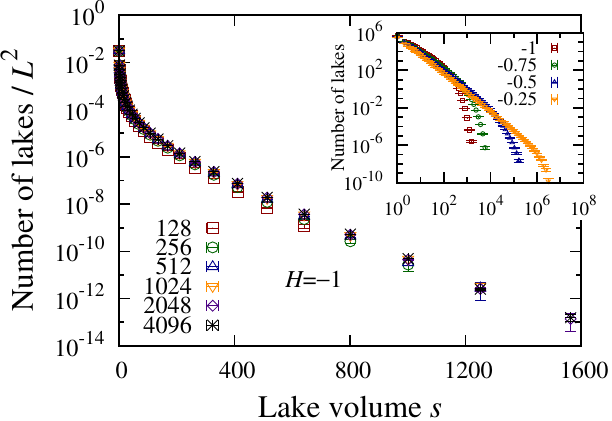}
	\caption{
		\label{fig:wr_corr_plot_1_uncorr_regime_and_evolution}
		(Color online)
		Uncorrelated regime and evolution with $H$.
		Main plot: Number of lakes of size $s$ per site as function of $s$ with ${H=-1}$ and for different lattice sizes $L$.
		Inset: Evolution of the lake size distribution with $H$, for fixed ${L=4096}$.
		The sites have height $0$ or $1$ with equal probability $1/2$.
		Results are averages over $10^4$ samples.
	}
\end{figure}
\begin{figure}
	\includegraphics[width=\columnwidth]{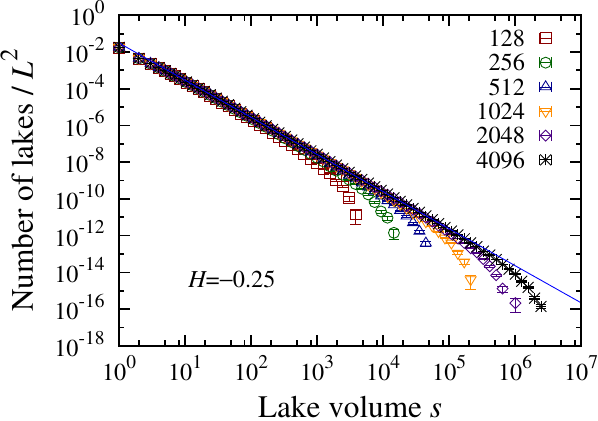}
	\caption{
		\label{fig:wr_corr_plot_2_corr_percolation_regime_and_slope_two}
		(Color online)
		Correlated percolation regime.
		Number of lakes per site for ${H=-0.25}$ and different lattice sizes $L$.
		The solid blue line is a guide to the eye with slope $-2.02$.
	}
\end{figure}
\begin{figure}
	\includegraphics[width=\columnwidth]{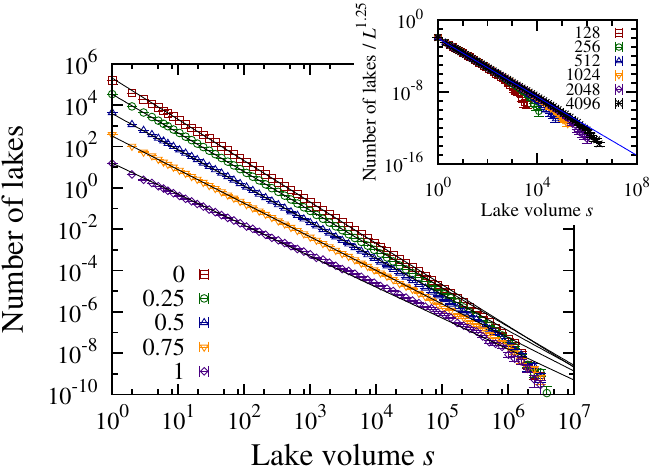}
	\caption{
		\label{fig:wr_corr_plot_3_non_sa_regime_exponent_predictions}
		(Color online)
		Main plot: Number of lakes as function of the lake volume for different non-negative $H$ and ${L=4096}$.
		Straight black lines are guides to the eye with slope ${2-H/2}$ \cite{Kondev95, Olami96, Kondev00}.
		Inset: Number of lakes, rescaled by ${L^{2-H}=L^{1.25}}$ \cite{Olami96}, as function of the lake volume, for ${H=0.75}$ and different lattice sizes $L$.
		The solid blue line is a guide to the eye with slope ${2-H/2 = 1.625}$ \cite{Olami96, Kondev95, Kondev00}.
	}
\end{figure}

Looking at the images in Figs.~\ref{fig::water_retention_chapter_continuum_512_H075} and \ref{fig::prob0_canonical_snaps}, one anticipates that, for correlated landscapes, the lake volumes vary considerably, while in the uncorrelated case, the lake sizes are more homogeneous.
To quantify this observation, we measure the number of lakes of volume $s$ as function of $s$.
Figure~\ref{fig:wr_corr_plot_1_uncorr_regime_and_evolution} shows the distribution of lake volumes for ${H=-1}$, corresponding to uncorrelated heights, and its inset compares data for different values of $H$.
One observes that, for negative $H$, the curves decay fast with increasing lake volume.
By contrast, for increasing $H$, the distributions display a power law regime; see Figs.~\ref{fig:wr_corr_plot_2_corr_percolation_regime_and_slope_two} and \ref{fig:wr_corr_plot_3_non_sa_regime_exponent_predictions}.
To understand the dependence of the volume distribution shape on $H$, we consider results for percolation on long-range correlated \cite{Weinrib84, Prakash92, Schmittbuhl93, Mandre11, Schrenk13b} and rough \cite{Schmittbuhl93, Kondev95, Olami96} surfaces.
In two dimensions, the generalized Harris criterion states that long-range correlations of the type considered here do not affect the nature of the percolation transition for ${H\leq-3/4}$ \cite{Weinrib83, Weinrib84, Schmittbuhl93, Sandler04, Schrenk13b}.
For the corresponding values of $H$, the lakes are sub-critical percolation clusters, as ${p=0.5}$ is significantly lower than $p_c$; the lake size distribution thus decays exponentially for large sizes \cite{Stauffer94}, as seen in Fig.~\ref{fig:wr_corr_plot_1_uncorr_regime_and_evolution}.
For ${-3/4\leq H \leq 0}$, the critical exponents are known to depend continuously on $H$, a phenomenon called \emph{correlated percolation} \cite{Weinrib84, Prakash92, Schmittbuhl93, Mandre11, Schrenk13b}.
In addition, the site percolation threshold $p_c$ decreases from ${p_c=0.59274602}$ \cite{Ziff11b} for the uncorrelated case, to ${p_c=0.5}$ as $H$ approaches zero \cite{Prakash92, Schrenk13b}.
In this range of $H$, we observe power laws with exponential cutoffs, see e.g. Fig.~\ref{fig:wr_corr_plot_2_corr_percolation_regime_and_slope_two}.
Finally, for ${H\geq0}$ it has been argued by Kondev \emph{et al.} \cite{Kondev95, Kondev00} and by Olami and Zeitak \cite{Olami96}, based on scaling arguments, that the cluster size distribution follows power laws, with the exponent depending on $H$.
As seen in Fig.~\ref{fig:wr_corr_plot_3_non_sa_regime_exponent_predictions}, this is consistent with our data, and in particular, the exponents of the power laws are consistent with the value ${2-H/2}$ predicted analytically in Refs.~\cite{Kondev95, Olami96, Kondev00}.
\section{\label{sec::fin_wr_chapter}Final remarks}
Concluding, we studied the water retention model \cite{Knecht11} on correlated and uncorrelated surfaces.
We confirmed some numerical results of Ref.~\cite{Knecht11} for the uncorrelated case and solved the model exactly for lattice size $L=3$.
It was found that long-range correlations decrease the retention capacity of random landscapes.
The decomposition of the retention for discrete landscapes [see Eq.~(\ref{eqn::retention_decomposition})] does also hold for the correlated case.
Here this intriguing result has been found numerically and we hope that it can be proven in the future.
For $H\geq0$, the lake-size distribution follows a power law, which can be quantitatively explained using the results of Kondev \emph{et al.} \cite{Kondev95, Kondev00} as well as Olami and Zeitak \cite{Olami96}.
In the future, it would be interesting to investigate different lattice geometries and boundary shapes.
In addition it could be possible to study the maximum, and the actual, water retention of real landscapes on earth \cite{Saberi13}.
\begin{acknowledgments}
We acknowledge financial support from the ETH Risk Center, the Brazilian institute INCT-SC, and (ERC) Advanced grant number FP7-319968-FlowCCS of the European Research Council.
K.J.S. acknowledges useful discussions with V.~H.~P.~Louzada and N.~Pos\'e.
\end{acknowledgments}
\appendix
\section{\label{sec::solution_test_L3_wr_chapter}Solution for small lattices}
\begin{figure}
	\includegraphics[width=0.5\columnwidth]{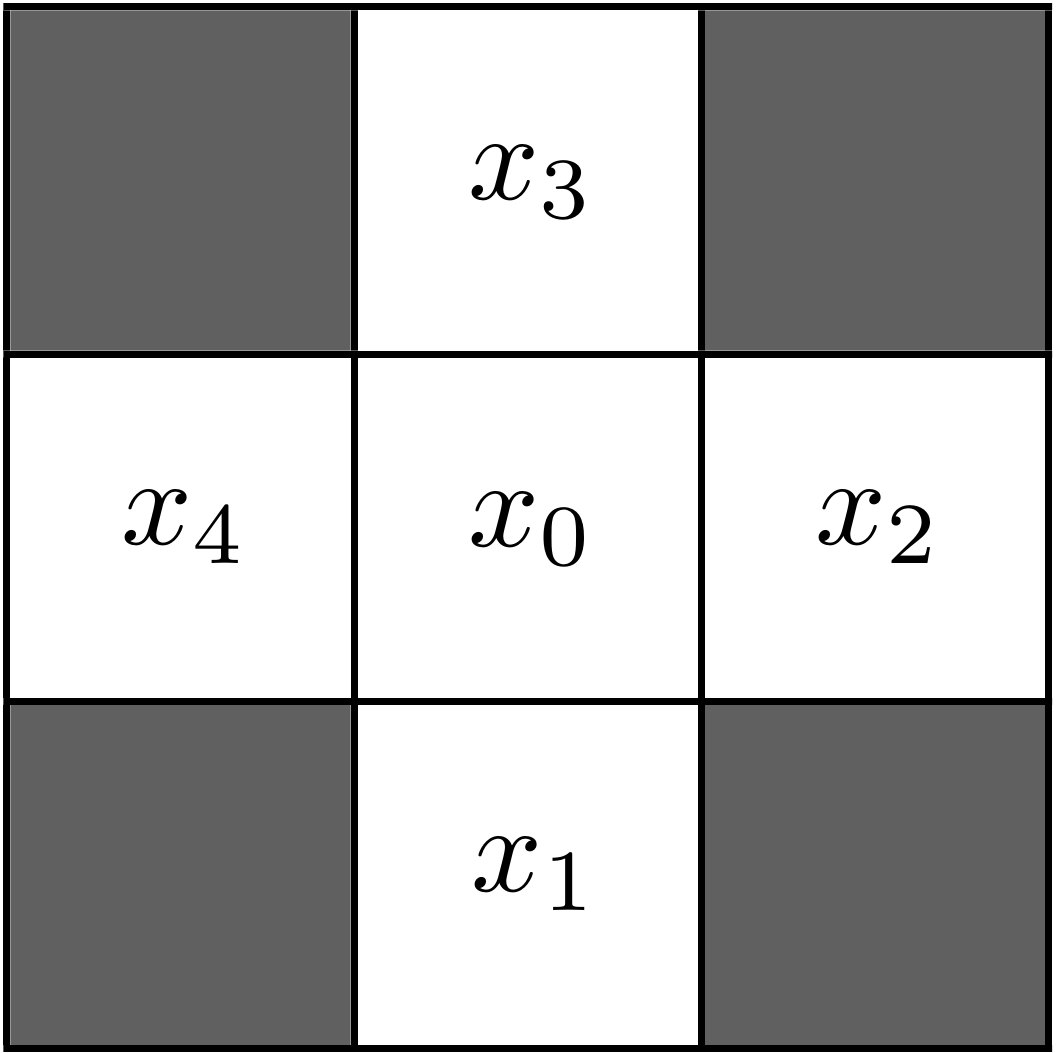}
	\caption{
		Illustration of the notation of the analysis on the ${L=3}$ square lattice.
		Note that the heights of the four corner sites (filled, gray) do not influence the volume retained by the lattice.
		\label{fig::retention_l3_sketch}
	}
\end{figure}
\begin{figure}
	\includegraphics[width=\columnwidth]{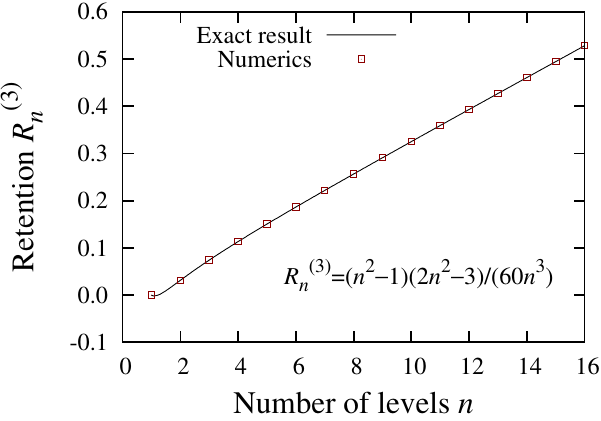}
	\caption{
		(Color online)
		Retention $R_n^{(3)}$ of an $n$-level system of size ${L=3}$ as function of $n$.
		The solid line is the exact result and the squares show the simulation data.
		The numerical results are averages over $10^7$ samples.
		\label{fig::wr_L3_n}
	}
\end{figure}
To confirm that the retention is monotonic in $n$ for small $L$ (see Fig.~\ref{fig::wr_FixedSSeveralL_PerSite}) and for further testing our simulation setup it is useful to compare the results with exactly solvable cases of the water retention model.
As suggested in Ref.~\cite{Knecht11}, we consider an $n$-level lattice of size ${L=3}$, see Fig.~\ref{fig::retention_l3_sketch}, and calculate $R_n^{(3)}$.
The four corner sites of the lattice are irrelevant because, due to the geometry of the square lattice, their heights do not influence the retained volume.
Let us call the height of the center site $x_0$.
If any of the four heights $x_1$, $x_2$, $x_3$, or $x_4$ is lower than or equal to $x_0$, the retained volume of the configuration is zero.
Otherwise, it is given by the difference between the lowest of the four relevant heights (where additional water would flow out to the border of the lattice) and the center height.
Therefore, for a given configuration, the retained volume $V$ is given by
\begin{equation}
	\label{eqn:vol_wr}
	\begin{array}{lll}
	V(x_0,x_1,x_2,x_3,x_4) &=& [\min\{x_1,x_2,x_3,x_4\}-x_0]\\& &\times\theta(x_1-x_0)\theta(x_2-x_0)\\& &\times\theta(x_3-x_0)\theta(x_4-x_0).
	\end{array}
\end{equation}
The heights of the lattice sites are independently and uniformly distributed integers in $[0,n-1]$, therefore we have to calculate the average
\begin{equation}
	R_n^{(3)} = \frac{1}{n^5} \sum_{x_0=0}^{n-1} \sum_{x_1=0}^{n-1} \sum_{x_2=0}^{n-1} \sum_{x_3=0}^{n-1} \sum_{x_4=0}^{n-1} V(x_0,x_1,x_2,x_3,x_4).
\end{equation}
Inserting the expression for the volume given in Eq.~(\ref{eqn:vol_wr}) yields (see Ref.~\cite{Knecht11}):
\begin{equation}
	\label{eqn::exact_l3_final}
	R_n^{(3)} = \frac{(n^2-1)(2n^2-3)}{60n^3}.
\end{equation}
Figure \ref{fig::wr_L3_n} shows the agreement between Eq.~(\ref{eqn::exact_l3_final}) and our simulation results.
The result of Eq.~(\ref{eqn::exact_l3_final}) is also consistent with the decomposition of the retention capacity in Eq.~(\ref{eqn::retention_decomposition}):
For two-level surfaces of size $3$ one has
\begin{equation}
	R_2^{(3)}(p) = p(1-p)^4.
\end{equation}
Inserting this into Eq.~(\ref{eqn::retention_decomposition}) recovers the result of Eq.~(\ref{eqn::exact_l3_final}).
\bibliography{bibliography}
\end{document}